\documentclass[a4paper,10pt,twoside]{cpc-hepnp}

\usepackage{multicol}
\usepackage{graphicx}
\usepackage{booktabs}
\usepackage{amssymb,bm,mathrsfs,bbm,amscd}
\usepackage[tbtags]{amsmath}
\usepackage{lastpage}

\begin{document}

\fancyhead[r]{\small Submitted to 'Chinese Physics C'}
\fancyfoot[C]{\small xxx-\thepage}

\title{Radiation field simulation of C-ADS Injector II\thanks{Supported by XDA03000000 }}

\author{%
      REN Guang-Yi$^{1,1)}$\email{gyren@mail.ustc.edu.cn}%
\quad ZENG Ming$^{2,2)}$\email{zengming@tsinghua.edu.cn}%
\quad HE Yuan$^{3}$\\
\quad LI Yu-Xiong$^{1}$
\quad LI Wei-Min$^{1}$
}
\maketitle

\address{%
$^1$ National Synchrotron Radiation Lab, University of Science and Technology of China, Hefei, 230029, China\\
$^2$ Tsinghua University, Beijing, 100084, China\\
$^3$ Institute of Modern Physics, Chinese Academy of Sciences, Lanzhou, 730000, China\\
}

\begin{abstract}
The injector of C-ADS (Chinese Accelerator Driven Sub-critical System) project is a high current, fully super-conducting proton accelerator. Meanwhile, a BLM system is indispensable to this facility, especially in low energy
segments. This paper presents some basic simulations for 10MeV proton by Monte Carlo program FLUKA, as well
as the distributions on different secondary particles in three
aspects: angular, energy spectrum and current. These results are beneficial to selecting the detector type and its
location and determining its dynamic range matching different requirements for both fast and slow beam loss. Further, the major impact of the background is analyzed, such as superconducting cavity X radiation and radiation caused by material activation, in this paper as well. This work is meaningful in BLM system research.
\end{abstract}

\begin{keyword}
C-ADS, FLUKA, BLM
\end{keyword}

\begin{pacs}
24.10.Lx, 29.27.Eg, 41.85.Qg
\end{pacs}

\footnotetext[0]{\hspace*{-3mm}\raisebox{0.3ex}{$\scriptstyle\copyright$}2013
Chinese Physical Society and the Institute of High Energy Physics
of the Chinese Academy of Sciences and the Institute
of Modern Physics of the Chinese Academy of Sciences and IOP Publishing Ltd}%

\begin{multicols}{2}

\section{Introduction}
\vspace{6mm}

The advantages of the Accelerator-Driven System (ADS) had been widely investigated, including the ability of transmuting High-Level radioactive Waste (HLW).
The Injector-\uppercase\expandafter{\romannumeral2}, as a possible front-end of C-ADS linear accelerator, consisting of an ECR ion source, a 2.1 MeV room-temperature RFQ at 162.5 MHz and superconducting half wave resonator (HWR) cavities at 162.5 MHz. The sc accelerating section will stimulate the proton from 2.1 MeV to 10 MeV. The Injector-\uppercase\expandafter{\romannumeral2} will operate in cw mode with a high average current of 10mA~\cite{background}.

Beam  loss  monitors (BLMs) are  common  devices  used  in hadron and lepton accelerators. Depending on accelerator specifics, BLMs could be just diagnostics or could play an essential  role  in  the  machine  protection  system  (MPS).
Beam loss control is one of the bottlenecks for beam power increasing for this high intensity machine.Different from the room-temperature accelerating section at the front-end of SNS or J-PARC, C-ADS will accelerate the proton by sc section from 2.1 MeV. Moreover, the sc cavity is more sensitive to beam loss than room-temperature accelerating section. Thus, it¡¯s necessary to provide a high level beam loss monitor system. In this paper, studies on radiation field caused by beam loss of C-ADS injector \uppercase\expandafter{\romannumeral2} would be fully discussed.

\section{Monte Carlo simulation}

\subsection{FLUKA Code}

The detailed analysis of radiation field of C-ADS injector \uppercase\expandafter{\romannumeral2} presented in this paper have been performed by the latest version of FLUKA Monte Carlo Code (version2011.2). FLUKA is a well benchmarked general purpose tool for calculations of particle transport and interactions with matters covering an extended range of applications, for example proton and electron accelerator shielding, target design, calorimetry, activation and dosimetry, cosmic ray studies, and radiotherapy~\cite{fluka}.

\subsection{Interaction between proton and niobium}

Protons lose their energy by ionization and atomic excitation while passing through matters such as the niobium, and the main formula governing this process is the Bethe-Bloch equation~\cite{sns}. While, the proton range and its secondary particles distribution are concerned about in this paper hereafter.
The average range of the 10 MeV proton in silicon is 667.4 $\mu$m ~\cite{protonrange}. The 10MeV proton range in niobium could be figured out according to the formula below:

\begin{displaymath}
    {R_i\over R_o}={{\rho_o \sqrt{A_i}}\over {\rho_i \sqrt{A_o}}}
\end{displaymath}

Where $\rho$ is the density; A means the atomic mass number; R represents the particle range. Consequently, the average range of the 10 MeV proton in niobium is 330.7 $\mu$m.
Then, 10 MeV proton pencil-beam bombarding the niobium target is simulated by FLUKA. The niobium target is a cylinder of 10 cm in diameter and 0.3 cm in thickness. The simulated range of proton is 312.3 $\mu$m, which approximates to the estimate based on the formula.

\begin{center}
\includegraphics[width=8cm]{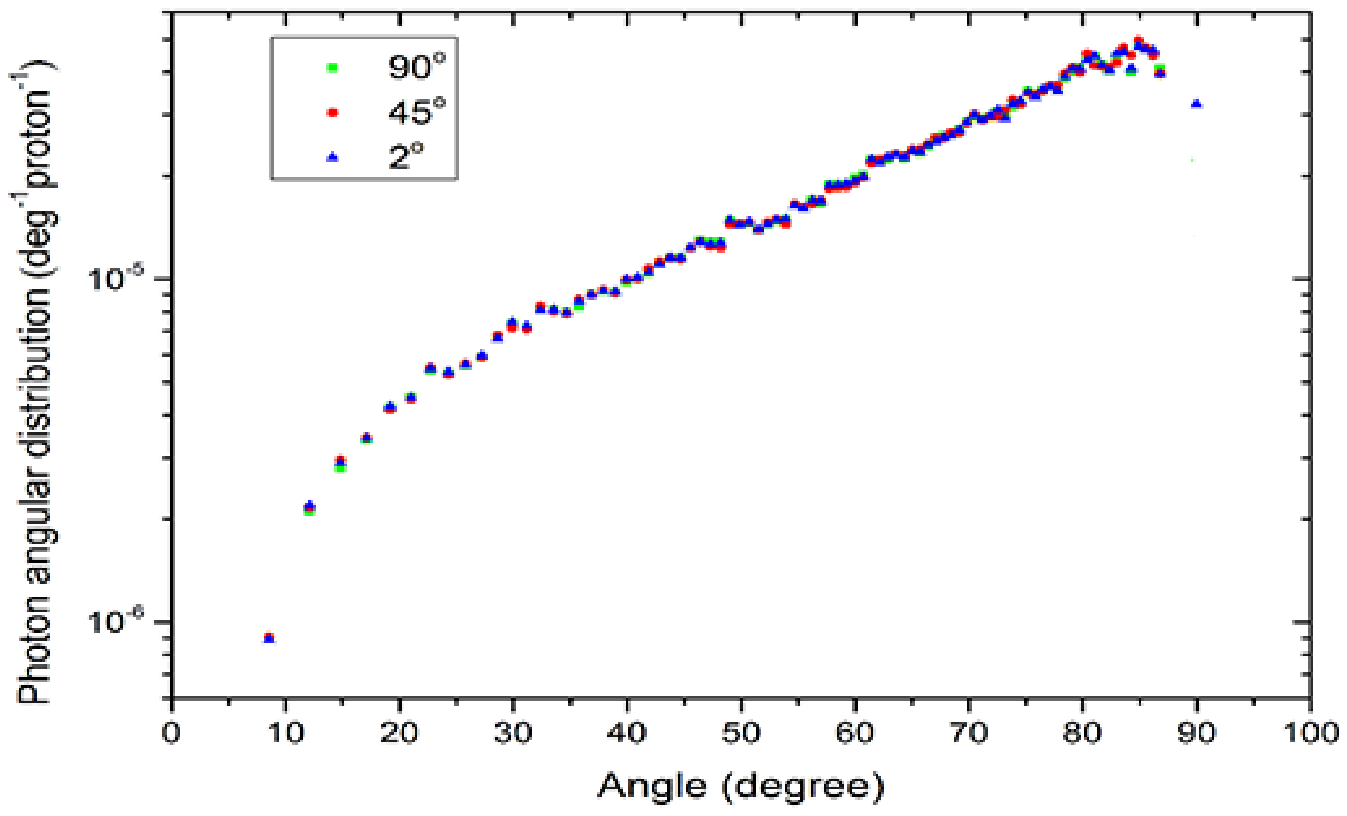}
\figcaption{\label{fig1}   Photon angular distribution in case of three different proton incidence angles. }
\end{center}

Fig.1 demonstrates the secondary photon angular distribution of different proton incidence angle. The angle is between the particle direction and the cylindrical bottom. The angular distributions for three incidence angles are simlar due to the short proton incident range. In conclusion, the distribution of secondary particles is irrelevant to the proton incidence angle when the proton energy is below 10MeV.

\begin{center}
\includegraphics[width=8cm]{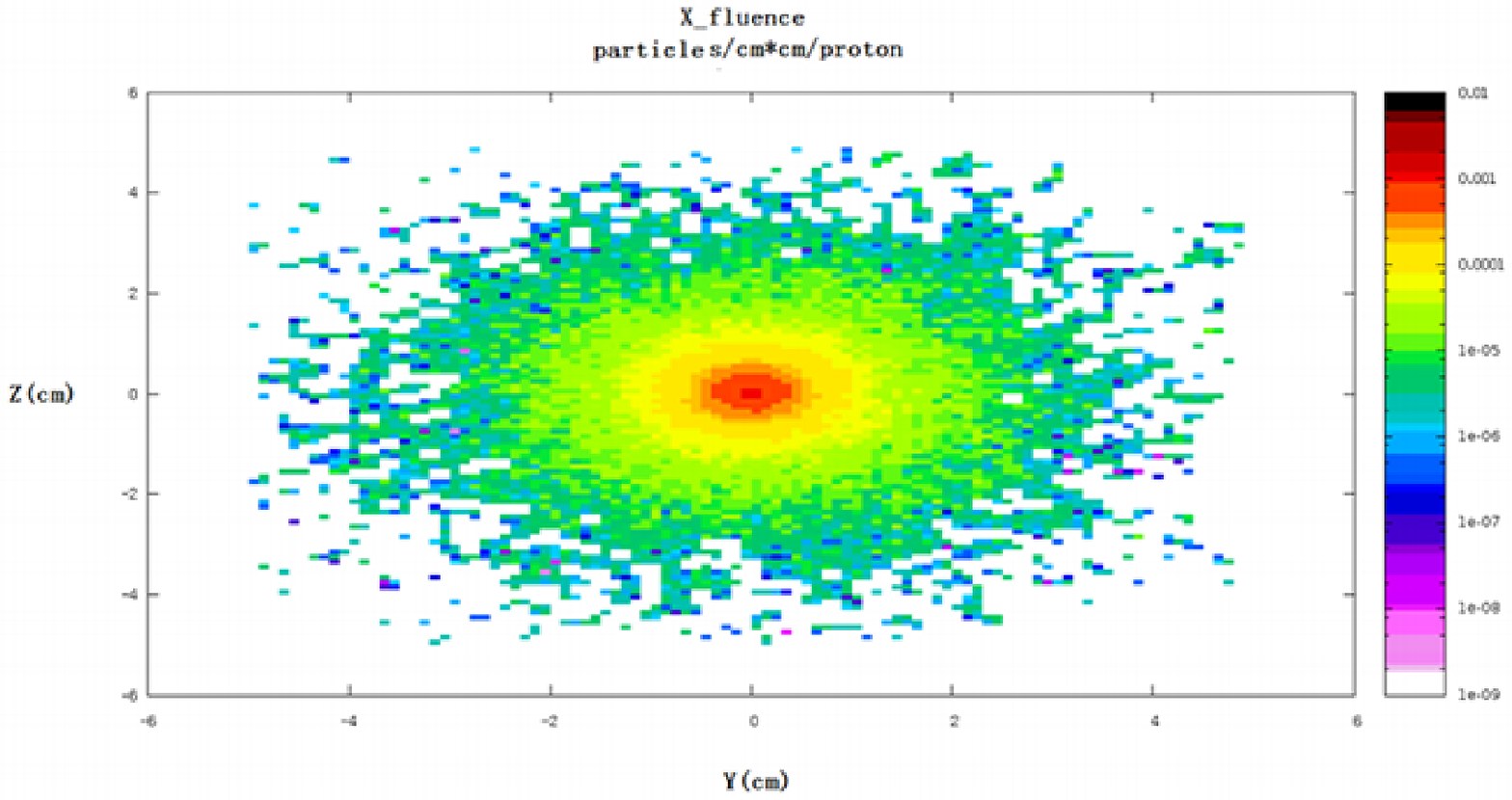}
\figcaption{\label{fig2}   Photon fluence of cylindrical bottom. }
\end{center}

The curve in Fig.1 reaches the summit around $45^{\circ}$ , and this phenomenon explains more exhibiting in Fig.2. According to Fig.2, the maximum fluence of photon performed in the center point. Besides, the fluence distribution of electron and neutron is similar to photon which is favorable to electron detect. Because of the very short electron range, electron could provide proton loss location. However, for photon or neutron, it's difficult for the detector to distinguish where they derived from. On the other side, photon or neutron detector will provide larger detection region.

\subsection{ Proton loss in sc cavity}

C-ADS Injector \uppercase\expandafter{\romannumeral2} accelerates the proton from 2.1 MeV to 10 MeV by sc cavity which has higher accelerating gradient than DTL or CCL and is more sensitive to beam loss for its superconducting state and difficult cleanliness requirements of cavity surface. This is the reason why the sc cavity demands concentrating from us. Because this research is attempting to explore the possibility of beam loss monitor for sc cavity, the 10MeV beam energy which will produce the most secondary particles is selected.

\begin{center}
\includegraphics[width=8cm]{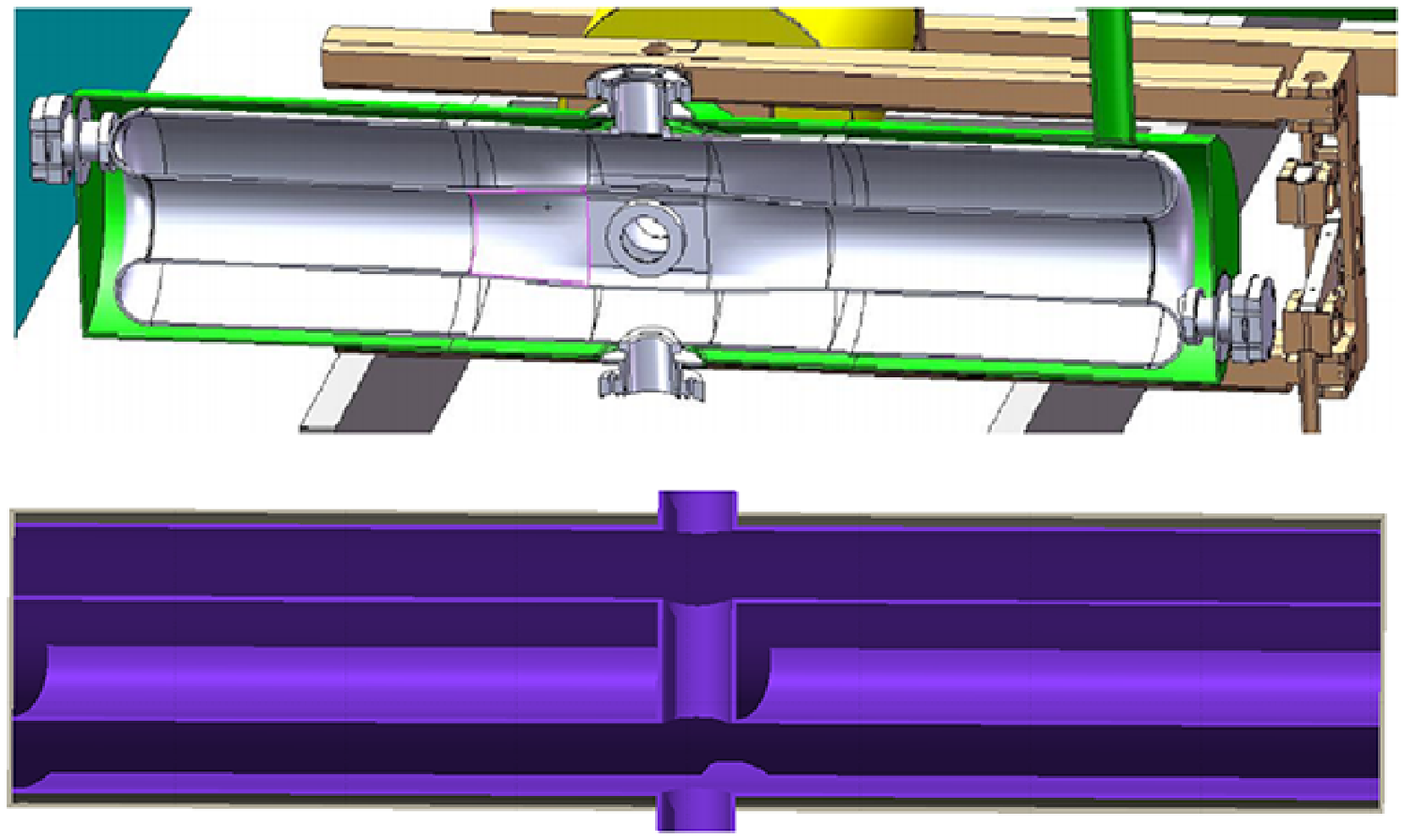}
\figcaption{\label{fig3}   Upside one is left-view of HWR cavity and the below is top-view of FLUKA model.}
\end{center}

Fig.3 depicts the sc HWR cavity and FLUKA simulation model. In the sc HWR cavity (the upper figure in Fig 3), the white structure is niobium cavity and the green one is titanium as insulation and support. The thickness of niobium and titanium are equally 3 mm. The protons pass through the sc cavity by the vacuum part in the center of HWR cavity, with liquid helium filling the gap between niobium and titanium to keep the low temperature.

Two representative beam loss points were selected and exhibited in Fig.4. If a detector is assigned on the right side of the cavity, the detected photon fluence of Fig.4 (b) will be two magnitude lower than Fig.4 (a). It¡¯s resemble to secondary electron and neutron which represents that monitoring the loss of situation (b) is more arduous than (a) until the detectors could perform rightly in the liquid helium or be placed in both side.

\begin{center}
\includegraphics[width=8cm]{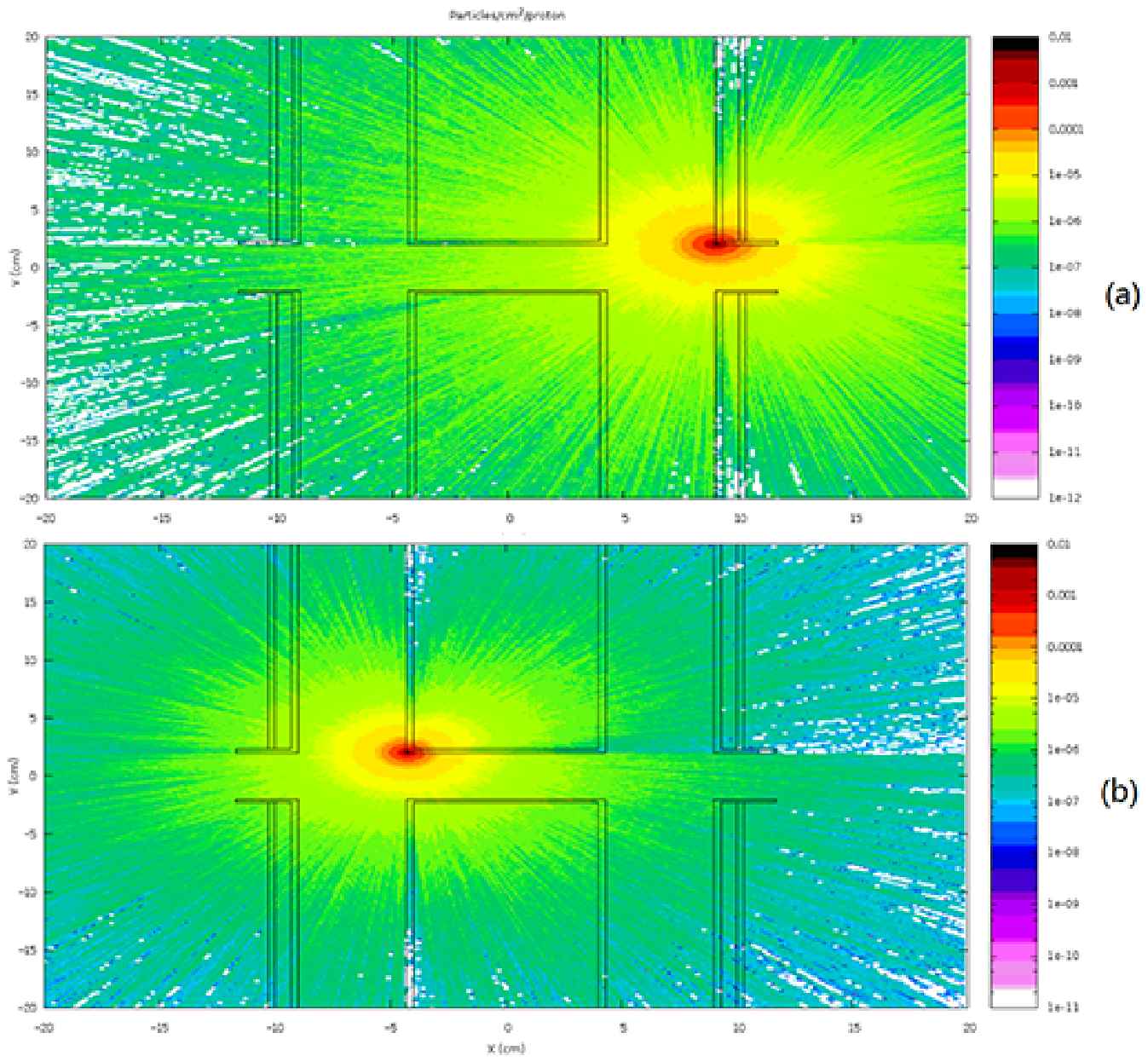}
\figcaption{\label{fig4}   Two representative beam loss points simulation. }
\end{center}

\subsection{Field Emission in sc cavity}

The sc accelerator may produce cavity X-ray when it operate on high accelerating gradient.Acknowledgedly, these cavity X-ray are caused by Field Emission (FE)~\cite{fieldelectron}. FE is due to quantum tunneling of electrons from microscopic defects on the RF surface, assisted by the cavity's electric field. Most field emitters occur in the high-electric-field regions near the cavity irises. In the process of accelerating the FE charges (electron mostly), the sped current impacts the cavity walls with diffracting bremsstrahlung x rays.

Since the accelerating voltage is 0.78 MV for the C-ADS sc HWR cavity, the energy of FE electron would be less than 0.78 MeV. If 0.78 MeV electrons hit the sc cavity at the point as show in Fig.4 (a), the FLUKA simulation result shows that a electron would produce some photon and secondary electron. The yield of photon is higher than the situation of proton loss. There is no neutron yielded.

\subsection{The simulation of delayed process}

Materials activation is inevitable for any particle accelerator. The radiation field of radionuclide should is treated seriously. For the simulation purposes, the normal loss which is homogeneously distributed along the HWR cavity inner surface should be taken into consideration. The accepted average beam loss limitation is 1W/m, representing that 1 $cm^{2}$ surface would suffer 1E+8 protons (10MeV) per second. As indicated by an irradiation simulation executed by FLUKA, beam current is 1E+8 protons/second after one-week-long irradiation.

\vspace{6mm}

\begin{center}
\tabcaption{ \label{tab1}  The list of radionuclide}
\footnotesize
\begin{tabular*}{80mm}{c@{\extracolsep{\fill}}ccc}
\toprule Z/A & Half-life	    &Decay models	     \\
\hline
Zr(90m)       & 809.2 ms	&IT        \\
Nb(93m)       &16.13 y	    &IT	            \\
Nb(94)     &2.03E+4 y	    &$\beta$- 	                   \\
Nb(94m)     &6.263 m	    &IT99.5\% $\beta$-0.5\%	                  \\
Mo(93)     &4.0E+3 y	    &EC	                   \\
Mo(93m)     &6.85 h         &IT99.88\% EC0.12\%	                  \\
\bottomrule
\end{tabular*}
\end{center}

The main short half-life radionuclides are Zr(90m), Nb(93m), and Mo(93m), and their activity would reach balancing in two days. Hereafter, the secondary particles could be detected outside the titanium, with some photons, a few electrons and no neutrons. These photons, whose energy are below 2.425MeV and have characteristic spectrum.

\section{The BLM system}

The ultimate goal of a Beam Loss Monitor system is to identify the loss level and, if possible, the loss¡¯s location and time structure. For C-ADS Injector \uppercase\expandafter{\romannumeral2}, nevertheless, the primary intention is to protect the machine. As a result, the BLM should detect the uncontrolled loss as soon as possible and shut down the accelerator immediately.

\subsection{Detecting target analysis}

\begin{center}
\tabcaption{ \label{tab2}  secondary particles peak yield out of sc cavity in 1 $\,cm^2$ (s=second, e=electron, p=proton)}
\footnotesize
\begin{tabular*}{80mm}{c@{\extracolsep{\fill}}ccc}
\toprule source & photon	    &neutron	     &electron\\
\hline
proton loss       & 3.37E-5 /p	&1.43e-5 /p	     &4.34e-7 /p        \\
FE electron       &2.26E-3 /e	    &non	         &8.42e-7 /e     \\
nuclide decay     &1.81E+3 /s	    &non	         &little           \\
\bottomrule
\end{tabular*}
\end{center}

\begin{center}
\includegraphics[width=8cm]{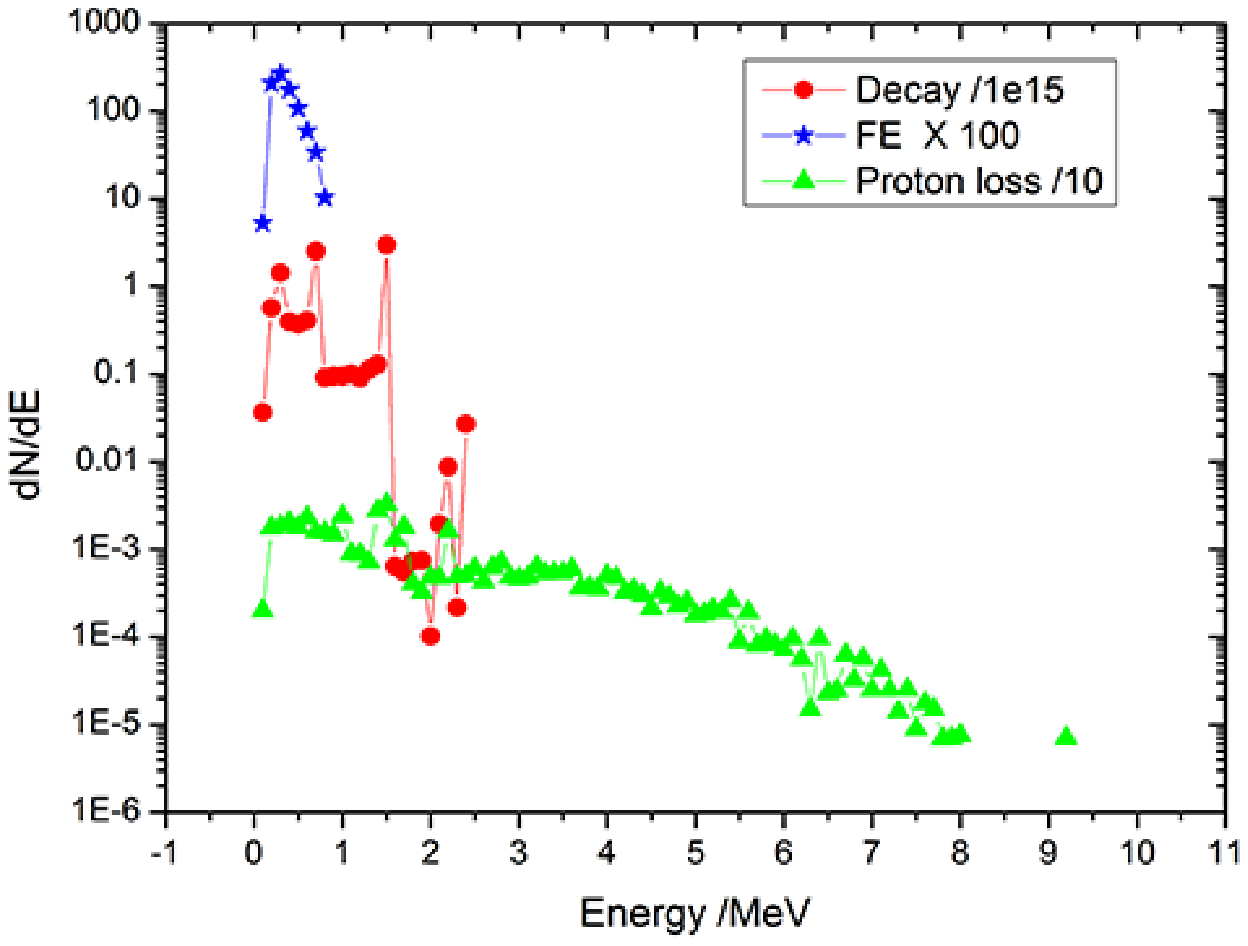}
\figcaption{\label{fig5}   Photon energy spectrum in three loss mechanism. }
\end{center}

Tab 2 and Fig. 5 give a complex relationship between the three main beam loss mechanisms and their secondary particles peak yield.  Neutron detectors are required because of the sole evidence for proton loss.  Meanwhile, the amount of photon yield is the largest, but acquiring the right signal is a challenge assignment to perform. For photon detection, the background would contain the photons caused by normal proton loss, $\gamma$-rays by residual radiation and X-rays by FE sometimes. As loss limitation is 1W/m, there will always be 5.18E+3 photon/second if the FE don¡¯t happen. When uncontrolled loss happens, beam collapses at somewhere. Micro-bunches with length of 6 ns (the specific condition with beam current 10 mA, frequency 162.5 MHz) will yield 3.8E+8 protons. One micro-bunch will produce 1.28E+4 photons, which is higher than the background without FE. It¡¯s divergent to estimate the influence of FE unless we shield the low energy photons as show in Fig.5.

\subsection{Detector type analysis}

According to the experience of SNS and J-PARC, neutron and photon are the main detecting particles. In J-PARC, scintillator and photomultiplier tube loss monitor is provided at low energy region for its fast time response ~\cite{jparc}. In SNS, Neutron Detector is useful for machine protection in MEBT and DTL while the ionization chamber is not sensitive enough at lower energy ~\cite{snsblm}.

Traditional BLM system for proton accelerators mainly consists of ionization chambers and
scintillation detectors such as neutron detectors. This combination is usually not sufficient to protect
low-energy high-power proton machines due to:\\
1) low radiation level from beam loss,\\
2) significant X-ray background near high-gradient superconducting RF cavities, and\\
3) poor loss localization with neutron detectors.

Because of the sc cavity, the selection of beam loss detector is more difficult for CADS injector. Under such circumstances, an ion chamber is not a good candidate to monitor the beam loss in the cold areas. It is also impossible to place scintillation detector in the cryomodule for its large size and low radiation hardness. Diamond detector has high radiation hardness and can be used at low temperature.
So diamond detectors is a promising candidate for BLMs of C-ADS injector \uppercase\expandafter{\romannumeral2}.

\section{Summary}

In this paper, the radiation field caused by beam loss of C-ADS injector \uppercase\expandafter{\romannumeral2} was simulated. The yield of secondary particle is small and the X-ray radiation produced by the RF or sc cavities is relatively big enough. Take the sc cavity of the CADS injector in to account, special consideration is needed for its BLM detector choosing and design, a combination of both neutron and gamma detection would be the solution, and the selection of the BLM detector location should be paid enough attention also.

\acknowledgments{The authors would like to express the most sincere thank to He Yuan (IMP) for his support of this work. }

\end{multicols}

\vspace{-1mm}
\centerline{\rule{80mm}{0.1pt}}
\vspace{2mm}

\begin{multicols}{2}

\end{multicols}

\clearpage

\end{document}